\documentclass[english,iop]{emulateapj}
\usepackage[T1]{fontenc}
\usepackage[latin9]{inputenc}
\usepackage{verbatim}
\usepackage{amssymb}
\usepackage{graphicx}

\makeatletter

\newcommand{\myemail}{lasota@iap.fr}
\def\spose#1{\hbox to 0pt{#1\hss}}
\def\lta{\mathrel{\spose{\lower 3pt\hbox{$\mathchar"218$}}
         \raise 2.0pt\hbox{$\mathchar"13C$}}}
\def\gta{\mathrel{\spose{\lower 3pt\hbox{$\mathchar"218$}}
         \raise 2.0pt\hbox{$\mathchar"13E$}}}
\newcommand{\Msun}{\mbox{$\rm M_{\odot}$}}

\slugcomment{Re-submitted to ApJ. (2011/04/16)}

\shorttitle{Variability of HLX-1}
\shortauthors{Lasota et al.}

\usepackage{babel}

\makeatother

\usepackage{babel}
\begin{document}
\global\long\def\Mo{\mathrm{M}_{\odot}}
 \global\long\def\Ro{\mathrm{R}_{\odot}}
 \global\long\def\Mbh{M_{\bullet}}
 \global\long\def\Ms{M_{\star}}
 \global\long\def\Rs{R_{\star}}
 \global\long\def\Mc{M_{c}}

\title{The origin of variability of the intermediate-mass black-hole ULX
system HLX-1 in ESO 243-49}

\author{J.-P. Lasota\altaffilmark{1,2},
T. Alexander\altaffilmark{3}, G. Dubus\altaffilmark{4,9}, D. Barret\altaffilmark{5,6}, S.A.
Farrell\altaffilmark{7,10}, N. Gehrels\altaffilmark{8}, O. Godet\altaffilmark{5,6}
\\
 and N.A. Webb\altaffilmark{5,6}}











\altaffiltext{1}{Institut d'Astrophysique de Paris, UMR 7095 CNRS,
UPMC Univ Paris 06, 98bis~Boulevard Arago, 75014 Paris, France\\
 Electronic address: {\myemail}} 
\altaffiltext{2}{Astronomical Observatory, Jagiellonian University,
ul. Orla 171, 30-244 Kraków, Poland}
\altaffiltext{3}{Department
of Particle Physics and Astrophysics, Faculty of Physics, Weizmann
Institute of Science, POB 26, Rehovot 76100, Israel}
\altaffiltext{4}{UJF-Grenoble
1/CNRS-INSU, Institut de Planétologie et d'Astrophysique de Grenoble
(IPAG) UMR 5274, Grenoble, F-38041, France} \altaffiltext{5}{Université
de Toulouse; Université Paul Sabatier - Observatoire Midi-Pyrénées,
Institut de Recherche en Astrophysique et Planétologie (IRAP), Toulouse,
France} \altaffiltext{6}{Centre National de la Recherche Scientifique;
IRAP; 9 Avenue du Colonel Roche, BP 44346, F-31028 Toulouse cedex
4, France } \altaffiltext{7}{Department of Physics and Astronomy,
University of Leicester, University Road, Leicester,~LE1 7RH, UK}
\altaffiltext{8}{NASA/Goddard Space Flight Center, Greenbelt,
MD 20771, USA} \altaffiltext{9}{also: chercheur associé at the
Institut d'Astrophysique de Paris} \altaffiltext{10}{at present:
Sydney Institute for Astronomy (SIfA), School of Physics, The University of Sydney, Sydney, NSW 2006, Australia}


\begin{abstract}
The ultra-luminous ($L_X \lta 10^{42}\, \rm erg/s$) intermediate-mass black-hole system HLX-1 in the
ESO 243-49 galaxy exhibits variability with a possible recurrence
time of a few hundred days. Finding the origin of this variability
would constrain the still largely unknown properties of this
extraordinary object. Since it exhibits a hardness-intensity behavior
characteristic of black-hole X-ray transients, we have analyzed
the variability of HLX-1 in the framework of the disk instability
model that explains outbursts of such systems. We find that the long-term
variability of HLX-1 is unlikely to be explained by a model in which
outbursts are triggered by thermal-viscous instabilities in an accretion
disc. Possible alternatives
include the instability in a radiation-pressure dominated disk but
we argue that a more likely explanation is a modulated mass-transfer
due to tidal stripping of a star in an eccentric orbit around
the intermediate-mass black hole.
We consider an evolutionary scenario leading to the creation of such
a system and estimate the probability of its observation. We conclude,
using a simplified dynamical model of the post-collapse cluster, that
no more than $1/100$ to $1/10$ of $\Mbh\mathrm{\lesssim10^{4}\,\Mo}$
IMBHs -- formed by run-away stellar mergers in the dense collapsed
cores of young clusters -- could have a $\mathrm{few\times}\,1\,\Mo$
Main-Sequence star evolve to an AGB on an orbit eccentric enough for mass transfer at periapse,
while avoiding collisional destruction or being scattered into the
IMBH by 2-body encounters. The finite but low probability of this configuration is consistent with the uniqueness of HLX-1.
We note, however, that the actual response of a standard accretion disk to bursts of mass transfer may be too slow to explain
the observations unless the orbit is close to parabolic (and hence even rarer) and/or additional heating, presumably
linked to the highly time-dependent gravitational potential, are invoked.
\end{abstract}


\section{Introduction}

\label{s:intro}

HLX-1 is the brightest ultra-luminous X-ray \citep[ULX; see][for a review]{roberts07}
source known, located in the outskirts of the edge-on S0a spiral galaxy
ESO 243-49 with a maximum luminosity of $\sim$10$^{42}$ erg s$^{-1}$
\citep{Farrell_nature,godetetal09}. The recent discovery \citep{redshift}
in the optical HLX-1 spectrum of an emission line consistent with
$H_{\alpha}$ at the redshift of ESO 243-49 ($z=0.0223$) irrevocably
confirms its association with this galaxy at a distance of 95 Mpc.
With observed X-ray luminosities reaching above $10^{42}$ erg\,s$^{-1}$
HLX-1 is super-Eddington if the black-hole's mass is less than $\sim10^{4}\,\Msun$.
Beaming effects \citep[e.g.][]{king08,koerding02} have been proposed
as viable mechanisms for producing the apparent super-Eddington luminosities
seen from other ULXs. However, beaming is unlikely to explain HLX-1's
extreme luminosity due to the observed large-scale variability (which
appears similar to that seen from Galactic stellar mass black hole
binaries that are not viewed down the jet-axis) and the luminosity
of the $H_{\alpha}$ line \citep[which is an order of magnitude above that expected from reprocessing in the local absorbing material;][]{redshift}.
Based on AGN-type scaling the $H_{\alpha}$ luminosity might suggest
a mass $\lta 1500\,\Msun$\,\citep{redshift}, but since it
is not clear how the line is related to the accreting system this
estimate is highly uncertain. By taking the conservative assumption
that HLX-1 exceeds the Eddington limit by no more than a factor
of 10 \citep{begelman02}, \citet{Farrell_nature} placed a lower
limit on the black hole mass of 500 $\Msun$. However, \citet{godetetal10}
obtained from a disc-blackbody fit a peak luminosity of $1.3\times10^{42}$
erg\,s$^{-1}$ (August 29, 2010) and a temperature $2.7\times10^{6}$\,K.
Comparing this value to that obtained from the (non-relativistic)
formula for the effective temperature at the inner edge of an accretion
disk around a black hole:
\begin{equation}
T_{{\rm in}}\approx6\times10^{6}\,\left(\frac{L_{42}}{\eta_{0.1}\, M_{4}^{2}}\right)^{1/4}\, x^{-3/4}\,{\rm K\,,\label{tin}}
\end{equation}
 where $L_{42}$ is the luminosity in units of $10^{42}$ erg\,s$^{-1}$,
$\eta=0.1\eta_{0.1}$ is the accretion efficiency ($\lta0.4$), $M_{4}$
the black hole mass in units of $10\,000\,\Msun$ and $x=c^{2}R/2GM$
is the radius measured in units of the Schwarzschild radius, one concludes
that $M\gta10^{4}\,\Msun$.
\begin{figure*}
\center \includegraphics[width=1\textwidth]{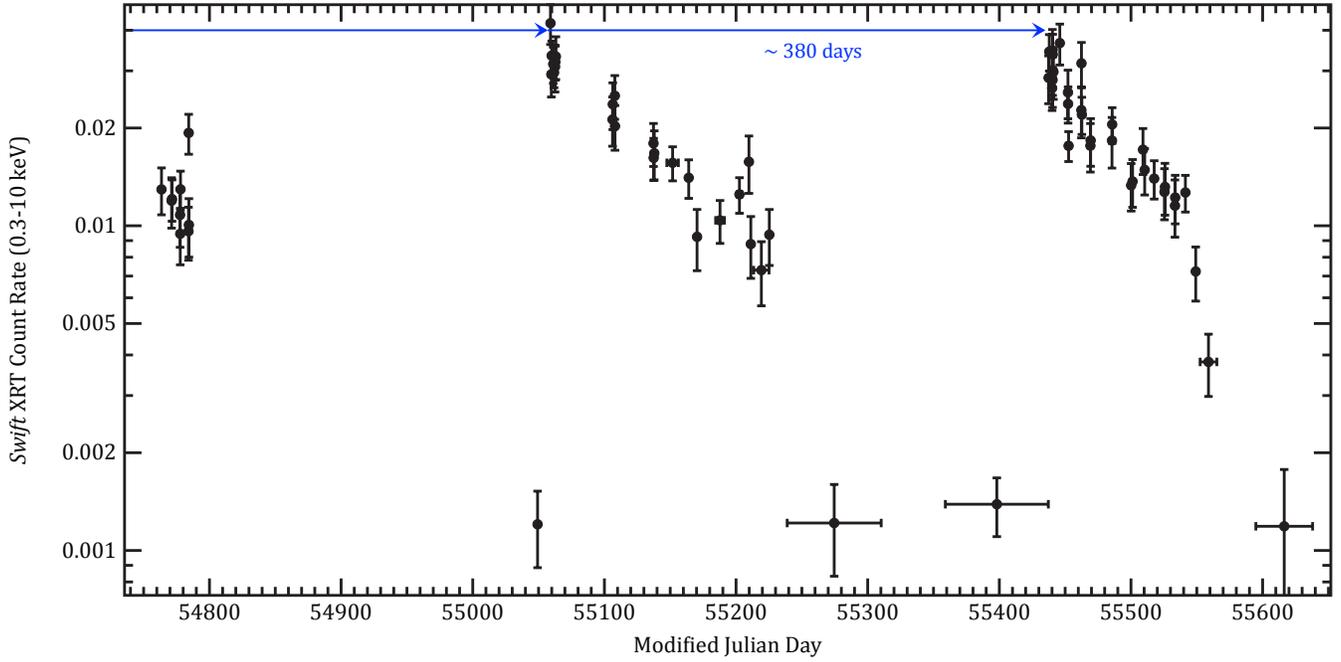}
\caption{The 2008 - 2011 \textit{Swift}-XRT Photon Counting grade 0-12 light-curve
of HLX-1 in the 0.3-10 keV energy range.}
\label{lcurve}
\end{figure*}

This conclusion is supported by more refined methods such as fitting
the {\em XMM-Newton} European Photon Imaging Camera (EPIC) spectra
with various models (\textsf{KerrBB, KerrD, slim-disc, BHSPEC}) which favor a mass
$M\sim 10^{4}\,\Msun$ \citep{godetetal11,davisetal}. The same is true
of the normalization of the \textsf{diskbb} model. One should add that the
temperature and luminosity observed during the previous luminosity
peak \citep[on the 2009-08-16;][]{godetetal09} were very close to those
determined recently.

After almost two years of monitoring with {\em Swift}, one sees
that the X-ray light-curve of HLX-1 shows variability with a characteristic
time of $\sim10^{7}$ s (Fig. \ref{lcurve}; \citet{godetetal10}).
The available data seem to indicate a FRED (Fast Rise Exponential
Decay) -- type shape. The recurrence time seems to be $\sim$ 380 days\footnote{373 days according to \citet{kong}}
\citep[consistent also with XMM1 and two Rosat observations, see][]{Webetal11}
and the decay-time from (the first) maximum is $\sim90$ days. The
rise is very steep and occurs over a timescale of about one week.
The amplitude of the X-ray variations is a factor of $\sim20\,-\,50$.
The decay-from-maximum light-curve shows small ``re-flares''
not dissimilar to those sometimes observed in low-mass X-ray transient
sources \citep[e.g.][]{chenetal}. The end of the second outburst's light-curve
is typical of X-ray transients suggesting the propagation of a cooling
front. In addition, the hardness of the X-ray spectrum of HLX-1 follows
the trend observed in X-transients: it increases with declining luminosity.
A recent deep observation with \emph{XMM-Newton} on 2010 May 14 (MJD
55330) confirms that the spectrum does harden at lower luminosities
($\sim$3 $\times$ 10$^{40}$ erg s$^{-1}$; Servillat et al. in
preparation). The analysis of these data suggests that there
may be some unresolved contribution from the nucleus of ESO 243-49
in the X-ray emission. However, spectral modeling (using a thermal
plasma model to represent emission from the galaxy) indicates that
this is likely to contribute no more than $\sim$20$\%$ of the total
observed flux, and thus that the low state luminosity is dominated
by emission from HLX-1.

These properties of the HLX-1 variability suggest that it could result
from the same thermal-viscous instability that drives the outbursts
of X-ray transient systems and dwarf-nova stars, and that the HLX-1
behaviour might be described by a variant of the corresponding disk
instability model \citep[DIM;
  see][for a review]{lasota01-1}.

\section{Accretion disk instabilities}

\label{s:instability}

\subsection{Accretion disk size according to the DIM}

The luminosity of HLX-1 ($\sim10^{42}$ erg s$^{-1}$) and the variability
timescale ($\sim10^{7}$ s) imply a huge accretion rate onto the compact
object, of the order of $10^{-4}$ M$_{\odot}$ yr$^{-1}$. Sustaining
such a high mass transfer rate excludes wind accretion.
We will therefore assume that the putative stellar companion of HLX-1
fills its Roche-lobe, if only during part of its orbit, losing matter
that forms an accretion disk around the black hole. This assumption
is also supported by the observed intensity related spectral changes.
The black hole mass is assumed to be around 10$^{4}$ M$_{\odot}$
so the mass ratio $q=\Mbh/\Ms$, where $\Mbh$ and $\Ms$ are respectively
the black-hole and stellar-companion masses, will typically be $q\approx10^{-4}-10^{-3}\ll1$
and the Roche-lobe radius can be written
\begin{equation}
R_{L}=a(q/3)^{1/3}\,,\label{roche}
\end{equation}
where $a$ is the orbital separation \citep[][for the moment we assume a circular orbit]{bp-77}.
When $q$ is so small, matter circularizes very close to the donor
star and the 2:1 Lindblad resonance appears at a radius $\approx0.63a$
within the primary's Roche lobe \citep{lin}, affecting the accretion
disk formation and structure \citep[see][for a discussion]{yungetal06}.
In the following we will assume the disk forms and has a size $R_{{\rm D}}\approx a$.

Here, we assume the variability in HLX-1 is related to the disk instability
model. This requires that the disk is big enough to allow its temperature
to fall below the temperature where hydrogen recombines \citep{lasota01-1}.
According to the DIM the accretion rate at outburst maximum is nearly
constant through the disk and roughly equal to the (upper) critical
accretion rate \citep{ldk}:
\begin{equation}
\dot{M}_{{\rm crit}}^{+}\approx3\times10^{-6}~\alpha^{-0.01}R_{13}^{2.64}~M_{4}^{-0.89}\,{\rm M_{\odot}{\rm yr}^{-1}}\,,\label{mdotcrit}
\end{equation}
where $R_{13}$ is the disk radius in units of $10^{13}$\,cm and
$\Mbh\,=\, M_{4}\,\times10^{4}\,{\rm M_{\odot}}$ is the mass of the
black hole. This limit is essentially independent of the Shakura-Sunyaev
disk-viscosity parameter $\alpha$. Therefore the maximum (bolometric)
luminosity is equal to
\begin{equation}
L_{{\rm max}}\approx1.6\times10^{40}\ \eta_{0.1}~R_{13}^{2.64}~M_{4}^{-0.89}\,{\rm erg\, s^{-1}\,.}\label{lmax}
\end{equation}
 Hence for $L_{{\rm max}}$ the disk radius will be equal to
\begin{equation}
R_{{\rm D}}^{{\rm max}}\approx5.2\times10^{13}\eta_{0.1}^{-0.38}M_{4}^{0.34}{\rm \, cm}\label{rmax}
\end{equation}
Strictly speaking, for (\textsl{inside-out}) outbursts starting in
the inner disk regions the radius in Eq. (\ref{mdotcrit}) corresponds
to the distance reached by the outside-propagating heating front.
This distance can be shorter than the actual disk outer radius. In
such case, the value given in Eq. (\ref{rmax}) is only a lower limit
for the disk radius.
Note also that this assumes a non-irradiated
disk. Taking into account irradiation of the outer regions of the
disk by X-rays from the inner regions always moves the critical
radius  further out \citep{dubetal-99}. Alternatively, an
irradiation-dominated disk reprocessing passively
${\cal C} \approx 0.1$ of the X-ray luminosity would still see
its temperature fall below 8000 K (hydrogen ionization) only for
radii greater than 2$\times 10^{13}$ cm, similar to the above estimate.
So, again, Eq. (\ref{rmax}) is only a lower limit on the disk radius.

From the Kepler's law $a$ is
\begin{equation}
a=6.3\times10^{12}\ M_{4}^{1/3}P_{d}^{2/3}\,{\rm cm\,,}
\end{equation}
 with $P_{d}$ being the orbital period (in days). Combining with
Eq.(\ref{roche}), the mean density of the companion $\bar{\rho}$
can be written in terms of the orbital period only
\begin{equation}
\bar{\rho}\approx0.057\, P_{d}^{-2}\,{\rm g\, cm^{-3}\,.}
\end{equation}
A disk size $R_{{\rm D}}\ga5\times10^{13}$ cm implies an orbital
period $\ga$ 23 days. The secondary star mean density for the hypothetical
companion of HLX-1 gives $10^{-4}\ {\rm g\, cm^{-3}}$ suggesting
a red giant or a massive supergiant. A priori, this is consistent
with several possible evolutionary scenarios for HLX-1 \citep[][but see Sect. 3]{Farrelletal10}.

\subsection{Outburst properties}

The presumed outburst of HLX-1 does not look like a {}``standard''
full-blown outburst of a black-hole transient low-mass X-ray binary.
In all such phenomena, after the outburst, the system declines (sometimes
after one or two re-bounds) to a quiescent, very-low luminosity state.
According to the DIM the (non-constant) accretion rate is then everywhere
lower than the critical value given by Eq. (\ref{mdotcrit}) (in fact
it is lower than the lower critical value $\dot{M}_{{\rm crit}}^{-}$).
Assuming that the disk terminates at the ISCO (Innermost Stable Circular
Orbit) this implies a luminosity of $<2\times10^{30}$ erg\ s$^{-1}$.
Even allowing for inner disk evaporation one can only increase this
luminosity by 4 or 5, say, orders of magnitude \citep{lny,menouetal,dubetal-01},
still well below the observed $\sim2.6\times10^{40}$ erg\ s$^{-1}$.

\subsection{Irradiation dominated disk}

The moderate amplitude of the outburst in HLX-1 suggests that only
part of the disk is involved {\em i.e.} that the DIM heating and
cooling fronts propagate only in a restricted domain of the disk.
Cooling fronts can be stalled when X-ray irradiation from the inner
regions of the disk prevents cooling in the outer regions. If the
disk irradiation is directly tied to the mass accretion rate onto
the compact object this does not prevent the disk from emptying but
the decay light curve falls down exponentially \citep{king,dubetal-01}.
More complicated behavior can arise if X-ray irradiation is not directly
tied to the inner mass accretion rate.

Here, irradiation by a hot supergiant star located just outside of
the disk can have an impact on its stability, independently of the
mass accretion rate. Such a possibility has been advocated by \citet{Revni02}
in the case of outbursts of the (presumably) super-Eddington source
V4641 Sgr. Such irradiation would be constant in time (or
at least accretion-independent) and its effects different from the
accretion self-irradiation. Constant irradiation of an unstable disk
typically leads to long term, low-amplitude modulation of the inner
mass accretion rate. An example calculation is shown in \citet{Dubus05}.
The cooling front propagates inward down to the radius where constant
irradiation keeps the disk hot. The region participating in the instability
is limited with little of the disk mass accreted at each cycle.

Although determining the exact timescale and amplitude of the modulation
requires numerical calculations, the typical outburst timescale will
be linked to the viscous timescale of the minimum unstable radius
(where the temperature must be $\approx10^{4}$ K)
\begin{equation}
t_{{\rm vis}}=\frac{R^{2}}{\nu}\approx115\ \alpha^{-1}T_{4}^{-1}R_{13}^{1/2}M_{4}^{1/2}{\rm years},\label{tvis}
\end{equation}
 $T_{4}$ is the disk temperature in units of 10$^{4}$ K and we have
used the Shakura-Sunyaev prescription for the kinematic viscosity
coefficient $\nu=\alpha c_{s}^{2}/\Omega_{k}$. For comparison, the
case shown in \citet{Dubus05} had a modulation period of about 0.5
years but had $R_{13}\approx0.01$ and $M_{4}\approx0.001$. The model
appears difficult to reconcile with the fast timescale of the variations
in HLX-1. The disk must be at least as large as required by Eq. (\ref{rmax})
(actually larger by a factor 2 when including irradiation heating)
in order to be unstable in the first place. In order to obtain small
amplitude outbursts, the radius to which the cooling front propagates
(Eq.~\ref{tvis}) cannot be orders-of-magnitude lower than the outer
disk radius. The modulation timescale will be too long for any reasonable
set of parameters. We conclude that the variability in HLX-1 is unlikely
to be related to the DIM in the sense that it is difficult to reconcile
the timescales with a disk large enough that its temperature at the
outer radius is below $\approx10^{4}$ K. Therefore the accretion
disk in HLX-1 is very likely to be hot and thermally stable.

\subsection{Radiation-pressure dominated disk}

We note that in considering the origin of HLX-1's variability one
should take into account that it is a near-Eddington X-ray source.
For $L/L_{{\rm Edd}}\gta0.01$ the standard Shakura-Sunyaev disk is
radiation-pressure dominated and opacities are due mainly to electron-scattering.
Such a disk is thermally and viscously unstable and the resulting
variability has been modeled producing FRED-like outbursts with amplitude
factors $\lta100$ \citep{TL,lp91,szusz01}, thus interesting
in the context of HLX-1. The unstable region is limited to the inner
$\sim100R_{g}$ from the black hole so it is generally short although
\citet{mayer} find a recurrence time of $\approx250$ years for the
case of a 10$^{6}$ M$_{\odot}$ black hole accreting at 10\% of the
Eddington rate. Contrary to the case of the dwarf-nova DIM, the physics
of the radiation-pressure thermal instability is rather controversial
\citep{hirose1,hirose2}. It remains to be seen whether the amplitude
and recurrence time of the variability in HLX-1 can be reproduced
by a radiation-pressure dominated disk around a 10$^{4}$ M$_{\odot}$
black hole \citep{zheng11,thermvis}.

\section{Orbital origin of the long-term variability}

We have
shown above that the disk-instability origin of the observed long term
variability is very unlikely. Here, prompted by the recurrent
nature of the X-ray flux variations, we consider whether the variability
could be orbital in origin.

\subsection{Orbital modulation}

Other ULX systems, such as X41.4+60 in M82 and NGC 5408 X-1 are variable
on timescales of 62 and 115 days respectively \citep{kaaret,stroh}.
These variations have been interpreted as reflecting orbital modulations.
As with those systems, the companion star of HLX-1 would have to be
a massive supergiant star ($\bar{\rho}\approx10^{-7}{\rm g\, cm^{-3}}$)
if the 380 day timescale is the orbital period. One mechanism that
would cause the X-ray modulation is absorption and scattering by the
companion's stellar wind. However, this explanation is unlikely for
HLX-1 for three reasons. First, the X-ray flux is modulated only by
a factor $\sim2$ in the case of the aforementioned ULX sources compared
to a factor 20 -- 50 in HLX-1. Second, their light-curves are quasi-sinusoidal
in contrast with the FRED-like shape of HLX-1. Finally, the observations
of HLX-1 do not show a significant change in the X-ray absorption
column density at different luminosity levels (but this is challenging
to constrain within the statistical quality of the X-ray data available).

Recently \citet{king11} proposed that ULXs in globular clusters could
be Ultra Compact X-ray Binaries (UCXBs) in which a neutron star accretes,
at slightly super-Eddington rates, matter lost by a Roche-lobe filling companion white dwarf. \citet{king11}
suggested that a version of this model might apply to HLX-1 but left the 380-day
variability unexplained. In such a scenario one could be tempted to
identify this variability with the so-called super-orbital period observed e.g.
in the UCXB 4U 1820-303, where a variability  170 days is observed in addition to
the orbital period of $\sim 11$ minutes is observed \citep[see][and references therein]{zdz07b}.
In this system the intrinsic luminosity varies by a factor of $\gta 2$ only
but larger amplitudes could be envisaged in the framework of a hierarchical triple
system model \citep{zdz07a}. However, the hardness-intensity relation in 4U 1820-303
shows a pattern completely different from that observed in HLX-1. Of course, as mentioned
by \citet{king11}, if the $H_\alpha$ line is emitted by the accretion flow the UCXB model
is ruled out. Clearly more observations in the optical domain are required to decide the viability of this (and
other) models.

\subsection{Modulated mass transfer from a donor in an eccentric orbit}

A scenario in which the disk in an eccentric binary would exist semi-permanently
(the viscous time of a standard accretion disk being long, see Eq.
(\ref{tvis})) and outbursts would be triggered by increased mass-transfer
rate during the passage of the companion appears to be the only serious
possibility left. It would be, in a sense, the equivalent of the so-called
mass-transfer instability model advocated previously for dwarf-nova
stars \citep{BaPr81}. In the case of HLX-1 one has to show that such
an binary system can be formed and survive long enough to be observed.
Here, we present the outline of such a scenario, which we explore
in more detail elsewhere \citep{ala+11}. One uncertainty is how
eccentric the orbit must be to allow fast enough accretion that can
reflect the orbital modulation. The simple assumptions we make here
are not expected to hold for very high eccentricity, but they can
provide an upper limit on the probability of an eccentric binary donor in HLX-1.

\subsubsection{The set-up}

For the moment observations provide only very general constraints
on possible evolutionary models of our hypothetical binary system.
As an example we assume here that the $P_{\star}\approx380\,\mathrm{d}$
period observed in the X-ray light curve of HLX-1 reflects the orbital
period of an evolved donor of mass $\Ms$ and radius $\Rs$, in an
eccentric orbit around an IMBH of $\Mbh\sim{\cal O}(10^{4}\,\Mo)$.
To feed the accretion (quasi)-periodically, the orbit must graze the
tidal disruption radius $r_{t}\simeq R_{\star}(\Mbh/\Ms)^{1/3}$ at
periapse, $r_{p}=a_{\star}(1-e_{\star})$, where $a_{\star}$ is the
orbital semi-major axis (sma), and $e_{\star}$ the eccentricity.
Therefore one obtains
\begin{equation}
a_{\star}\simeq22.1\,\mathrm{AU}\,\left(\frac{P_{\star}}{380\,\mathrm{d}}\right)^{2/3}\left(\frac{M_{\bullet}}{\Mo}\right)^{1/3}\,,
\end{equation}
 ($22.1\,\mathbb{\mathrm{AU}}=7\times10^{14}\,\mathrm{cm}$). The
extremely high mass-loss requirement of ${\cal O}(10^{-4}\,\Mo\mathrm{yr^{-1}})$
strongly suggests an AGB donor \citep[initial mass range of $\sim 0.5\, \Mo$
to $10\, \Mo$;][]{habingolofsson}, which can reach $\sim10^{-5}-10^{-4}\,\Mo\,\mathrm{yr^{-1}}$ mass-loss
rate even without {}``tidal inducement'' \citep{bowenwilson}. We
will adopt here as a fiducial AGB progenitor a $M_{\star}=4\,\Mo$
star with a main-sequence (MS) radius of $2.3\,\Ro$ and therefore
a MS lifetime of $t_{\star}\simeq200\,\mathrm{Myr}$, which expands
to $R_{\star}\sim100\,\Ro$ in the AGB phase. The eccentricity required
to graze the tidal radius is then
\begin{eqnarray}
1-e_{t} & \!=\! & r_{t}/a_{\star}\!\nonumber \\
 & \simeq & 0.3\!\left(\frac{\Rs}{100\,\Ro}\right)\left(\frac{\Ms}{4\,\Mo}\right)^{-1/3}\left(\frac{P_{\star}}{380\,\mathrm{day}}\right)^{-2/3}\,.
\end{eqnarray}
 While still on the MS, the donor orbit is stable to both tidal and
gravitational wave decays, since $r_{p}\sim75r_{t}\sim10^{5}r_{g}$.
However, once it reaches the AGB phase, it will be quickly destroyed
within $10^{4}-10^{5}\,\mathrm{yr}$, due to a combination of the
high mass loss rate and the tidally driven orbital decay.

The scenario proposed here is different from the one suggested by
\citet{Hopmanetal04} to explain the non-periodic ULX in the young
cluster MGG-11 in M82. There, it was assumed that the progenitor was
tidally captured while still on the MS, and that the orbit decayed
and circularized by the tidal interaction before the star reached
the AGB, resulting in a relatively steady mass transfer rate on a
very short-period circular orbit evolving through emission of gravitational
radiation.

\subsubsection{Evolution and survival}

\label{sss:evol}

Run-away stellar mergers in the dense collapsed cores of young
clusters are widely considered to be a natural pathway to the formation
of an IMBH \citep{ebi+01,por+02b,gur+04,fre+06b,fre+06c}. The scenario
considered here is based on the simulations by \citet[][hereafter GFR04]{gur+04}%
\footnote{A scenario in which a nucleated dwarf galaxy is undergoing a period
of accretion due to a recent passage through the host galaxy is often
mentioned in the context of HLX-1. Although the presence of IMBH in
nuclei of dwarf galaxies has still to be demonstrated \citep[see e.g.][]{reinesetal11}
we think we are justified in considering this possibility just as
a slightly different realization of the scenario we are studying.%
}. They find that for a wide range of initial parameters, a dense cluster
with a broad Initial Mass Function (IMF) will undergo mass segregation and core collapse after
a time $t_{CC}\sim0.1t_{r,1/2}(0)$, where $t_{r,1/2}(0)$ is the
initial 2-body relaxation at the half-mass radius of the cluster.
When the core-collapse time is shorter than the minimal MS lifetime
for massive stars
\begin{equation}
t_{CC}<\min t_{\star}\sim3-4\,\mathrm{Myr},
\end{equation}
 the massive stars that segregate in the dense collapsed core can
undergo run-away mergers on a timescale shorter than the stellar evolutionary
timescale. If the merged {}``super-star'' can cool fast enough and
avoid rapid mass loss, it may collapse to form an IMBH.

Guided by the GFR04 simulations, in what follows we will assume that
this process forms an IMBH with a mass ratio $M_{\bullet}/M_{c}=0.002$,
where $M_{c}$ is the total initial mass of the cluster. As reference
we will use Model 2 of GFR04 in which a Salpeter IMF extending from
$0.2$--$120\, M_{\odot}$ is assumed (mean mass $\left\langle M_{\star}\right\rangle =0.7\, M_{\odot}$,
$\max M_{\star}/\left\langle M_{\star}\right\rangle =174$). The model
assumes a Plummer distribution with a length-scale $R_{c}=(2^{2/3}-1)^{1/2}r_{1/2}\simeq0.766r_{1/2}$,
where $r_{1/2}(0)$ is the initial half mass radius. The core-collapse
time is:
\begin{equation}
t_{CC}\approx4.7\,\frac{N_{6}}{\log10^{4}N_{6}}\frac{1}{\sqrt{\rho_{9}}}\,\mathrm{Myr\,}\label{e:tr12}
\end{equation}
 where $\rho_{0}=(10^{9}\, M_{\odot}\,\mathrm{pc^{-3}})\rho_{9}$
is the initial central density and $N_{c}=10^{6}N_{6}$ is the stellar
number. The condition $t_{CC}<3\,\mathrm{Myr}$ then translates to
$\rho_{0}\gtrsim10^{9}\,\Mo\,\mathrm{pc^{-3}}$ (for $\Mc=5\times10^{6}\,\Mo$,
$N_{c}=7.25\times10^{6}$, $R_{c}=0.11$ pc). Although young clusters
and super star clusters with relaxation times well below $30$ Myr
are observed \citep{fig+02,ho+96} it is less clear whether such extreme
central densities are realized in nature. A lower IMBH mass will ease
this problem since according to Eq. (\ref{e:tr12}), $\rho_{0}\sim M_{\bullet}^{2}$.

In such dense systems mass segregation proceeds very rapidly and accelerates
the evolution toward core collapse. Following the results of GFR04
(their Fig. 5) we take for the stars closest to the newly formed IMBH
$\left\langle M_{\star}\right\rangle _{h}=4\, M_{\odot}$, with a
corresponding radius $\left\langle \Rs\right\rangle _{h}=2.34\,\Ro$,
as the mean stellar mass and radius within the IMBH radius of influence,
$r_{h}=G\Mbh/\sigma_{c}^{2}$, where $\sigma_{c}^{2}$ is the typical
1D velocity dispersion of the host cluster ($\sigma_{c}^{2}(0)=GM_{c}/6R_{c}\simeq185\,\mathrm{km\, s^{-1}}$
for the Plummer model here). It is noteworthy that mass segregation
initially concentrates intermediate mass AGB progenitors near the
IMBH.

Following the collapse the dynamical response of the cluster by mass
segregation and expansion leads to the formation of a relaxed stellar
cusp inside $r_{h}$, with $n_{H}(r)\propto r^{-\alpha_{H}}$ ($\alpha_{H}=7/4$--$5/2$)
for the massive stars, and $n_{L}(r)\propto r^{-3/2}$ for the low-mass
stars \citep{bah+77,ale+09,kes+09,pre+10}. The cusp extends inward
down to the collision radius, $r_{in}=\max(r_{t},r_{\mathrm{coll}})=r_{\mathrm{coll}}\sim(\Mbh/\Ms)\Rs$,
at which 2-body relaxation ceases to be effective because velocities
are so high that only physical collisions can substantially change
the stellar orbits \citep{fra+76}. For a standard MS mass-radius
relation this gives
\begin{equation}
r_{\mathrm{coll}}\sim48.7(M_{\bullet}/10^{4}\, M_{\odot})(M_{\star}/M_{\odot})^{-0.43}\,\mathrm{AU},
\end{equation}
 so that for the cusp to extend down to $a_{\star}\simeq22$ AU the
stars must be more massive than $\sim3.7\, M_{\odot}$. One notices
that this lower limit coincides with the mass range predicted by pre-IMBH
mass segregation, while being low enough to include AGB progenitors.

Equilibrium is established when the flux of gravitational binding
energy that is released when stars are destroyed by the IMBH equals
the flux carried by the expanded cluster core \citep{heg+07}. Assuming
a Plummer initial distribution one obtains the mass of the cusp $M_{h}$
as \citep[e.g.][]{baumgetal04}
\begin{equation}
M_{h}\simeq[648/(3-\alpha_{H})](\Mbh/M_{c})^{2}M_{\bullet}\,.\label{e:Mh}
\end{equation}
In contrast to the radius of influence of a supermassive black hole
(SMBH) in an approximately isothermal galactic nucleus, that of an
IMBH contains only a very small number of stars, $N_{h}=M_{h}/\left\langle \Ms\right\rangle _{h}\sim{\cal O}(10)$.

Equations (\ref{e:tr12}) and (\ref{e:Mh}), together with the system
parameters $\Mbh$, $\Mbh/M_{c}$, $\alpha_{H},$ $\left\langle \Ms\right\rangle $
and ($\left\langle \Ms\right\rangle _{h}$,$\left\langle \Rs\right\rangle _{h}$)
fully describe the IMBH cusp structure and properties in this simplified
model, and allow calculating the mean number of stars on orbits with
$a\le a_{\star}$ and $e\ge e_{\star}$, $\left\langle N_{a,e}\right\rangle =N_{h}[(5/4)a_{\star}/r_{h}]^{3-\alpha_{H}}(1-e_{\star}^{2})$,
where an isotropic cusp is assumed. The Poisson probability for having
at least one star on a donor orbit is then $P_{1}=1-\exp[-\left\langle N_{a,e}\right\rangle ]$.

The cusp mass and its stellar density rise with the IMBH mass (Eq.
\ref{e:Mh}). As a consequence, $P_{1}$ increases with $\Mbh$, but
with it also the rate of destructive stellar collisions, and the drain
rate (star-star scatterings into the IMBH, \citealt{ale+04}). Over
the relevant range of IMBH masses, $P_{1}$ rises from $\lesssim10^{-3}$
to $\gtrsim0.1$. The IMBH mass that maximizes $P_{1}$ subject to
both the collisional and drain constraints lies in the range of $\mathrm{few}\times10^{3}\,\Mo$,
with $\max P_{1}\sim\mathrm{few\times0.01}$.

The observed residual UV fluxes from HLX-1, after subtracting a disk
model and correcting for extinction ($E_{B-V}=0.042$), are $4.8\times10^{-15}\,\mathrm{erg\, cm^{-2}\, s^{-1}}$
($L_{NUV}=5.2\times10^{39}\,\mathrm{erg\, s^{-1}}$ for isotropic
emission at $D=95\,\mathrm{Mpc}$) in the NUV ($2147\,\text{\AA}\,-\, 3467\,{\text{\AA}}$),
and $4.7\times10^{-15}\,\mathrm{erg\, cm^{-2}\, s^{-1}}$ ($L_{FUV}=5.1\times10^{39}\mathrm{erg\, s^{-1}}$
) in the FUV ($1233\,\text{\AA}\,-\, 1821\,\text{\AA}$) (based on a preliminary
analysis of HST data, Farrell et al., in preparation). These place
constraints on the properties of the hypothesized birth cluster of
the IMBH.

Stellar population synthesis models together with model atmospheres
allow to predict the UV flux as function of the IMF, metallicity and
age of the system. Generally, the older the cluster, the less UV it
emits. An older cluster directly implies a lower mass AGB progenitor,
that can spend longer on the MS. Preliminary modeling, both assuming
black body spectra, and using detailed stellar atmosphere models (the
\textsc{stars} code, \citealt{ste98}) indicates that a minimal cluster
age of $\sim(0.3-0.6)\,\mathrm{Gyr}$ yr is required for the cluster's
UV luminosity to fall below the residuals (Assuming a cluster mass
of $5\times10^{6}\,\Mo$ and solar metallicity). This corresponds
to an AGB progenitor of $\sim(2.7-3.5)\,\Mo$. Such longer-lived progenitors
are more susceptible to collisional destruction and scattering into
the IMBH, but are still probable at the $P_{1}\sim\mathrm{few\times0.01}$
level. The constraints on the AGB progenitor mass could be relaxed
somewhat by assuming a higher value of $\Mbh/M_{c}$, and possibly
by assuming a different metallicity. This requires a more systematic
study.

\begin{figure}
\begin{centering}
\includegraphics[width=1\columnwidth]{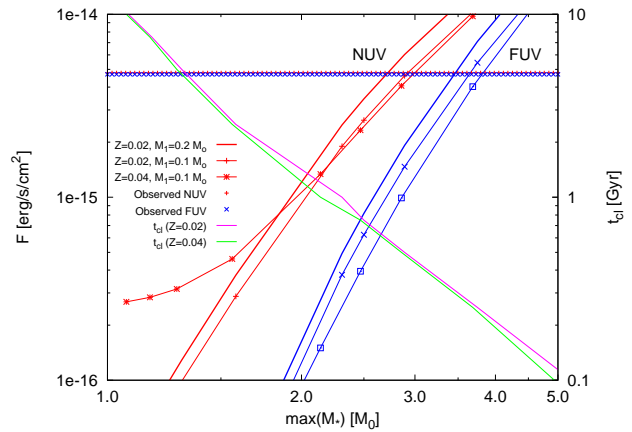}
\par\end{centering}

\caption{\label{f:UV}A comparison of the observed de-reddened limits of the
residual UV fluxes (after subtraction of a disk model) to stellar
population synthesis model predictions for different metallicities
($Z$) and low mass cutoffs ($M_{1}$) and assuming a $5\times10^{6}\,\Mo$
cluster, as function of the maximal progenitor mass (the turn-off
mass), which sets the age of the cluster (also shown) and hence its
minimal UV luminosity. }
\end{figure}

After mass transfer starts in the AGB phase, 2-body
perturbations from other stars can be neglected, since the relaxation
time at $r_h$ is $t_h \sim \mathrm{few} \times 10^6$ yr, and the
timescale to significantly affect the orbital eccentricity,
$(1-e_\star^2) t_h$, is longer than the $\sim 10^4$ yr maximal lifetime
of the donor, for all relevant values of the eccentricity.

Thus, while an IMBH with $\Ms\sim10^{4}\,\Mo$ is not excluded in
the context of the eccentric donor scenario, observations and theoretical
considerations favor a somewhat less massive IMBH with $\Mbh\sim\mathrm{few}\times10^{3}\,\Mbh$,
still consistent with the constraints derived from the observed accretion
emission (see Sec. \ref{s:intro}).

To summarize, we used a simplified model, based on results from $N$-body
simulations of runaway merger formation of a seed IMBH, to predict
the stellar distribution around a newly formed IMBH. We estimated
the probability of finding an AGB progenitor on an orbit that could
explain the long period variability of HLX-1 in terms of mass transfer
from an eccentric evolved donor. We find that between $1/100$ to
$1/10$ of IMBHs with $\Mbh\mathrm{\lesssim10^{4}\,\Mo}$ could have
a $\mathrm{few\times}\,1\,\Mo$ MS star evolve to an AGB while avoiding
collisional destruction or being scattered into the IMBH by 2-body
encounters. An important caveat is that the validity of the model
and our conclusions are limited by the neglect of the longer-term
evolution of the cusp and host cluster, and in particular by the neglect
of the effects of 2-body relaxation and fast resonant relaxation \citep{rau+96,hop+06a}
on the distribution of the stars around the IMBH.

The low probabilities are fully consistent with the apparent uniqueness of HLX-1.
Theoretically we can only make statements about the conditional probability since
we do not know the space density IMBH clusters but observationally, despite large-scale searches
(in Chandra data by \cite{liu11} and in XMM data by \citet{walton10}), there have not been
any other HLX-1 like objects found. Most probably HLX-1 is alone in the local Universe.

\subsubsection{Variability timescales}

\label{sss:ts}

In our scenario involving a $1-e=0.3$ binary in which a star
circles a $\sim10^{4}\Mo$ black hole on an 380 day orbit, the periapse
will be at $\sim10^{14}$\,cm. The mass loss and disk formation processes
is usually described by the star filling an {}``instantaneous''
Roche lobe at periapse \citep[see e.g.][in the context of high mass X-ray binaries]{Sep07,Lajoie11}.
This description is increasingly inaccurate as the eccentricity grows
but matter lost during the periapse passage is expected to circularize
at about this distance.
The transferred material will have to diffuse inwards, on a viscous
timescale. From the discussion in Sec.\ref{s:instability}, the disk
is most likely hot with a temperature $\geq10,000$\,K at the outer
edge. The diffusion timescale for the transferred material would be
hundreds of years rather than days according to Eq. \ref{tvis}. As
a result, the accretion timescale will smooth out the bursts in mass
transfer.

One way out of this difficulty is to increase the disk temperature.
Irradiation is unlikely to increase the disk temperature much: the
irradiation temperature $T_{{\rm irr}}=({\cal C}L_{X}/4\pi\sigma_{{\rm SB}}R^{2})^{1/4}$
is $\approx3500$\, K at $10^{14}$ cm (where ${\cal C}\approx10^{-3}$
parametrizes our ignorance of the irradiation geometry and albedo,
\citealt{dubetal-99}). Incoming material may shock heat the outer
disk to high temperatures, especially in the case of a tidal disruption
on a parabolic orbit when the material ejected from the star has a
large range of velocities relative to the Keplerian disk velocity,
up to the escape speed of the star \citep[see e.g.][]{rees}. The response
of the disk to a burst of mass transfer in this situation has not
been modeled. Note, however, that in close low-mass binary systems
the {}``hot spot'' resulting from the mass-transfer stream impact
significantly heats the outer disk without drastically affecting the
dynamical properties of the disk because the thermal timescale on
which the extra heating is radiated away in a thin disk is much shorter
than the viscous timescale \citep{buat,smak02}.

Another way to reduce the diffusion timescale is to reduce the
periapse distance. The viscous timescale is already down to 3 years
if periapse is at $r_{t}\sim10^{12}$\,cm (assuming the temperature
varies as $T\sim R^{-1/2}$, in Eq. \ref{tvis}). In such a model
the disk does not disappear completely in between mass transfer episodes
(in accord with the non-zero minimum flux) and the impulsive increase
in mass transfer leads to increase in lightcurve on timescales more
like a fraction of $t_{vis}$ ($\lta0.1$) and then decays on $t_{vis}$
or so. The passage of the star at periapse may also lead to the excitation
of waves in the disk that will enhance angular momentum transport
\citep{spruit}. Tidal waves may provide an additional source of heating
in the disk. The price to pay is that such a binary has a much
lower probability because (1) the orbit must be nearly parabolic,
$e\rightarrow1$, (2) the depletion of the phase space density of
orbits near the loss cone. which was not taken into account in our
simple estimate (sec. \ref{sss:evol}), and (3) the unstable nature
of such an orbit. In fact, this instability could lead to an observed
complete disruption within a few orbits (few years).

\section{Discussion and Conclusions}

The extremely high luminosity, light curve shape and X-ray spectrum
evolution of HLX-1 point toward disk accretion around a 10$^{4}$
M$_{\odot}$ black hole fueled by a Roche-lobe filling star. We have
examined the conditions under which the X-ray variability might be
explained by the disk instability model. We find this requires an
accretion disk much too large for corresponding timescales to be compatible
with the observed X-ray variability. Any accretion disk around HLX-1
is most likely small enough to be hot and stable against the DIM.
One cannot exclude that the variability is due to the instability
that can arise in radiation-pressure dominated disks. However, the
physics behind this instability is not well known and it is not clear
whether this will lead to the correct outburst amplitude and timescales.

The variability would be much easier to explain with a stellar-mass
black hole but the luminosity would obviously be problematic. Conversely,
the luminosity is no issue for a supermassive black hole but the amplitude
and timescales would be an insurmountable problem \citep{nags}.

We have shown, however, that a viable description of the HLX-1 variability
can be provided by a model in which enhanced mass transfer into a
quasi-permanent accretion disk is triggered by the passage at periapse
of an evolved (AGB) star circling the IMBH on an eccentric orbit.
Using a simplified model based on the results of N-body simulations
we concluded that such systems, although not common, are realistically
observable. However, the actual response of a standard
accretion disk to bursts of mass transfer may be too slow to explain
the observations unless the orbit is close to parabolic and/or additional
heating, presumably linked to the highly time-dependent gravitational
potential, is invoked.

The validity of our conclusions is limited by the neglect of the longer-term
evolution of the cusp and host cluster, in particular by neglecting
the effects of 2-body relaxation and fast resonant relaxation on the
distribution of the stars around the IMBH. In general, the very small
number of stars in the IMBH cusp casts doubts on the applicability
of the statistical approaches commonly used to analyze dynamics around
SMBHs. In addition, the comparatively high density of unbound (cluster)
stars in the cusp complicates the analogy with known results from
SMBH cusp dynamics. Progress in the analysis of the post-formation
evolution of IMBHs will require time-dependent modeling and
N-body simulations.

The further evolution of the intriguing variability pattern of HLX-1
(periodicity, amplitude) as well as observations of its optical counterpart
should shed light on the origin and nature of this extraordinary
system.

\acknowledgements{}

JPL thanks Ramesh Narayan, Andrzej Zdziarski and Richard Mushotsky
for inspiring discussions. We thank the referee whose report stimulated
additional work on the modulated transfer model. This work was supported
by the French Space Agency CNES and in part by the Polish MNiSW grant
N N203 380336 (JPL), the United Kingdom's STFC and the Australian
Research Council (SAF) and the European Commission via contracts ERC-StG-200911
(GD) and ERC-StG-202996 (TA).

\expandafter\ifx\csname natexlab\endcsname\relax\global\long\def\natexlab#1{#1}
 \fi

\end{document}